\documentclass[fleqn,usenatbib]{mnras}

\usepackage{newtxtext,newtxmath}

\usepackage[T1]{fontenc}
\usepackage{ae,aecompl}
\usepackage[dvipsnames]{xcolor}

\usepackage{graphicx}	
\usepackage{amsmath}	
\usepackage{amssymb}	

\title[Number of satellites vs. B/T]{A correlation between the number of satellites and the baryonic bulge-to-total ratio extending beyond the Local Group}

\author[Javanmardi \& Kroupa]{
Behnam Javanmardi$^{1}$\thanks{E-mail: behnam.javanmardi@obspm.fr}
and
Pavel Kroupa$^{2,3}$
\\
$^{1}$LESIA, Paris Observatory, PSL Univ., CNRS, Sorbonne Univ., Univ. Paris Diderot, 5 Place Jules Janssen, 92195 Meudon, France\\
$^{2}$Helmholtz Institut f\"ur Strahlen- und Kernphysik (HISKP), University of Bonn, Nussalle 14-16, 53121 Bonn, Germany\\
$^{3}$Charles University, Faculty of Mathematics and Physics, Astronomical Institute, V Hole\^sovivck\'ach 2, 180 00 Praha 8, Czech Republic
}

\date{}

\pubyear{2019}

\begin{document}
\label{firstpage}
\pagerange{\pageref{firstpage}--\pageref{lastpage}}
\maketitle

\begin{abstract}
Recent observations of the fields surrounding a few Milky-Way-like galaxies in the local Universe have become deep enough to enable investigations of the predictions of the standard $\Lambda$CDM cosmological model down to small scales outside the Local Group. Motivated by an observed correlation between the number of dwarf satellites ($N_{\rm sat}$) and the bulge-to-total baryonic mass ratios ($B/T$) of the three main galaxies in the Local Group, i.e. the Milky Way, Andromeda, and Triangulum (M33),  we use published data of three well-studied galaxies outside the Local Group, namely M81, Centaurus A, and M101, and their confirmed satellites, and we find a strong and significant correlation between $N_{\rm sat}$ and $B/T$. This presents itself in contradiction with the hitherto published results from cosmological simulations reporting an absence of a correlation between $N_{\rm sat}$ and $B/T$ in the $\Lambda$CDM model. We conclude that, based on the current data, the $N_{\rm sat}$ vs. $B/T$ correlation is no longer a property confined to only the Local Group.
\end{abstract}




\section{Introduction}
Until recently, the halos of only the Milky Way (MW) and Andromeda (M31) galaxies have been observed deep enough to enable studying different properties of their dwarf satellites \citep[e.g.][]{Grebel1997,McConnachie2012,Collins2015} and to test the expectations of galaxy formation and evolution scenarios down to small scales. The observations of these two galaxies, and the Local Group (LG) in general, has led to some of the long lasting challenges for the standard Lambda-Cold-Dark-Matter ($\Lambda$CDM) cosmological model, viz: the \textit{missing satellites} \citep{Klypin1999}, the \textit{too-big-to-fail} \citep{Boylan-Kolchin2011}, and the \textit{disk of satellites} \citep{Kroupa2005,Pawlowski2013} problems. The numerous observational, theoretical, and computational studies aiming at solving these issues have not reached a consensus so far \citep[see e.g.][for a review]{Bullock2017}. These problems required systematic and deep surveys for finding more dwarf satellites outside the LG. In the recent years, and thanks to various dedicated deep observations \citep[e.g.][]{Geha2017,Greco2018}, other galaxy groups are also turning into very useful laboratories for investigating the small scale predictions\footnote{We note though that the disk-of-satellites problem is not a small scale problem as it refers to the distribution of baryonic matter on scales of hundreds of kpc about major host galaxies.} of the $\Lambda$CDM model, and for learning about galaxy formation and evolution in general.

Another interesting feature of the LG that motivated dedicated studies \citep[such as ][]{Javanmardi2016,Lopez2016,Henkel2017} is an observed correlation between the number of dwarf satellites, $N_{\rm sat}$, and bulge mass ($M_{\rm bulge}$) of M33, MW, and M31 \citep{kroupa10}. This correlation appeared to be a puzzling one due to the following two reasons: i) in the $\Lambda$CDM model, $N_{\rm sat}$ is expected to correlate with only the mass of the dark matter halo, hence with the rotation velocity, of the disk galaxy, and ii) galaxies with similar rotation velocities \citep[e.g. M31 and M101,][]{Faber1979} but with very different $M_{\rm bulge}$ are observed in the Universe. Therefore, unlike the expected correlation between $N_{\rm sat}$ and rotation velocity, a correlation between $N_{\rm sat}$ and $M_{\rm bulge}$ was not expected.

To quantify the expectations from the $\Lambda$CDM model for this correlation, \citet{Javanmardi19} used data from the Millennium-II simulation \citep{Boylan-Kolchin2009} with the semi-analytic galaxy formation model of \citet{Guo2011} and reported three main results: i) for a sample of disk galaxies with a wide mass range, a weak correlation (with a large scatter) between $N_{\rm sat}$ and $M_{\rm bulge}$ emerges which reflects the seemingly independent correlations between these two quantities and the mass of the disk galaxy, ii) for disk galaxies with similar masses or rotation velocities, no correlation between $N_{\rm sat}$ and $M_{\rm bulge}$ is found, and iii) when normalizing $M_{\rm bulge}$ with the mass of the disk galaxy, i.e. when considering $B/T$ ratios, no correlation is found between $N_{\rm sat}$ and $B/T$ of disk galaxies with stellar mass $M_{\star}\approx10^{10}$-$10^{11} M_{\odot}$ in the $\Lambda$CDM model.

 Recently, observations of the galaxies M81, Centaurus A (Cen A), M94, and M101 (at distances of about 3.7, 3.8, 4.2, and 7 Mpc, respectively) have become deep enough to enable studying their satellite properties and their relation to their hosts \citep{Chiboucas_2013,Cronjevic2019,Muller2019,Smercina2018,Carlsten2019, Bennet2019}. In this contribution, we use the published data from these galaxy groups, together with the available data from M33, MW, and M31, to investigate the $N_{\rm sat}$ vs. $B/T$ correlation.\\

\section{Data and Analysis}\label{sec:data}
In this section, we first introduce the data we use in our analysis from different studies published in the literature. See Table \ref{tab:data} for a summary. We then apply a simple correlation analysis and present the results.
\subsection{Number of dwarf satellites}\label{sec:nsat} 
The information about the satellite populations of the galaxies in our study are as follows:
\begin{itemize}
    \item The LG galaxies: we refer to \citet{McConnachie2012} and \citet{McConnachie_2018}, for the information about MW and M31 satellites, respectively\footnote{Note that a possible, but not confirmed, satellite of M33 is reported by \citet{Martin2009} with $M_{\rm V}=-6.5$.}.
    
    \item M81: using the Canada-France-Hawaii Telescope (CFHT) and Hubble Space Telescope (HST) observations, \citet{Chiboucas2009} and \citet{Chiboucas_2013} have discovered and confirmed 14 new dwarf galaxies in the M81 group. Their survey covers at least $\approx$300 kpc projected distance around M81 \citep[and in some directions more than twice that, see figure 27 of][]{Chiboucas_2013}. They increased the number of known members of this group to 36 with its faintest dwarf having an r-band absolute magnitude of $M_{\rm r}$=-6.8 \citep[$M_{\rm V}\approx -8$,][]{Chiboucas_2013}. See their Table 4 for a full list of information. It is worth noting that we do not include two tidal dwarf galaxy candidates of M81 \citep[namely A0952 and Garlnd,][]{Karachentsev2002, Makarova2002} in our analysis due to uncertainties on their nature. We also note that we treat M82 as a satellite galaxy of M81 (rather than considering it separately) because M82 (being less than 100 kpc in projection away from M81) shows clear signs of disturbance and is undergoing a star-burst, indicating strong interactions between the two galaxies\footnote{We note that M33 shows no signs of interaction, i.e. is close to being nearly isolated, and therefore we consider it as a separate galaxy and not as a satellite of Andromeda.}.
    
     \item Cen A: \citet{Cronjevic2019} and \citet{Muller2019} have used the Panoramic Imaging Survey of Centaurus and Sculptor (PISCeS), the HST, and the ESO Very Large Telescope (VLT) to confirm the dwarf satellite candidates of Cen A, most of which were identified by \citet{Cronjevic2014}, \citet{Cronjevic2016}, \citet{Muller2015}, and \citet{Muller2017}. Together, \citet{Cronjevic2019} and \citet{Muller2019} report in total 31 confirmed dwarf satellites for Cen A with the faintest having a V-band absolute magnitude of $M_{\rm V}=-7.8$ mag. According to \citet{Muller2019}, the satellite population of Cen A is complete within its 200 kpc projected distance, while two of its already confirmed satellites, namely KK 211 and ESO 325-011, are at projected distances of 240 and 246 kpc, respectively \citep{Cronjevic2019}.
    
    \item M94: using deep Hyper Suprime-Cam (HSC) observations, \citet{Smercina2018} performed a survey of the field of M94 down to $M_{\rm V}=-9.1$ mag finding only two satellites for this galaxy. Regrettably, their survey covers only out to 150 kpc projected distance and while we discuss this galaxy in our study, we do not include it in our correlation analysis.
    
    \item M101: \citet{Danieli2017} and \citet{Bennet2019} used deep HST observations and measured the Tip of the Red Giant Branch (TRGB) distances, and \citet{Carlsten2019} have used the CFHT data and measured surface brightness fluctuations distances to satellite candidates around M101. These studies confirmed in total 8 satellites for this galaxy with the faintest having a V-band absolute magnitude of $M_{\rm V}=-8.2$ mag. Most of these candidates have been found in a wide survey using also the CFHT by \citet{Bennet2017} and some of them by \citet{Merritt2014}, \citet{Javanmardi2016}, and \citet{Mueller2017}. These surveys cover almost isotropically the 260 kpc virial radius of M101 \citep[see figure 1 in][]{Carlsten2019}.
\end{itemize}

The latter galaxy group imposes a magnitude limit on our analysis and to be able to compare the satellite populations of all galaxies, we only count the satellites with $M_{\rm V} \leq -8.2$ mag. In addition, we will perform our correlation analysis (Section \ref{sec:correlation}) with two projected distance limits; first considering only the satellites within 200 kpc projected distance to their hosts, and second increasing the projected distance limit to 250 kpc by assuming that the satellite population of Cen A is complete-enough out to that distance, see the last columns of Table \ref{tab:data} where $N^{200}_{sat}$ and $N^{250}_{sat}$ represent the number of satellites within 200 and 250 kpc of each galaxy, respectively.

\begin{figure*}
    \centering
    \includegraphics[width=\textwidth]{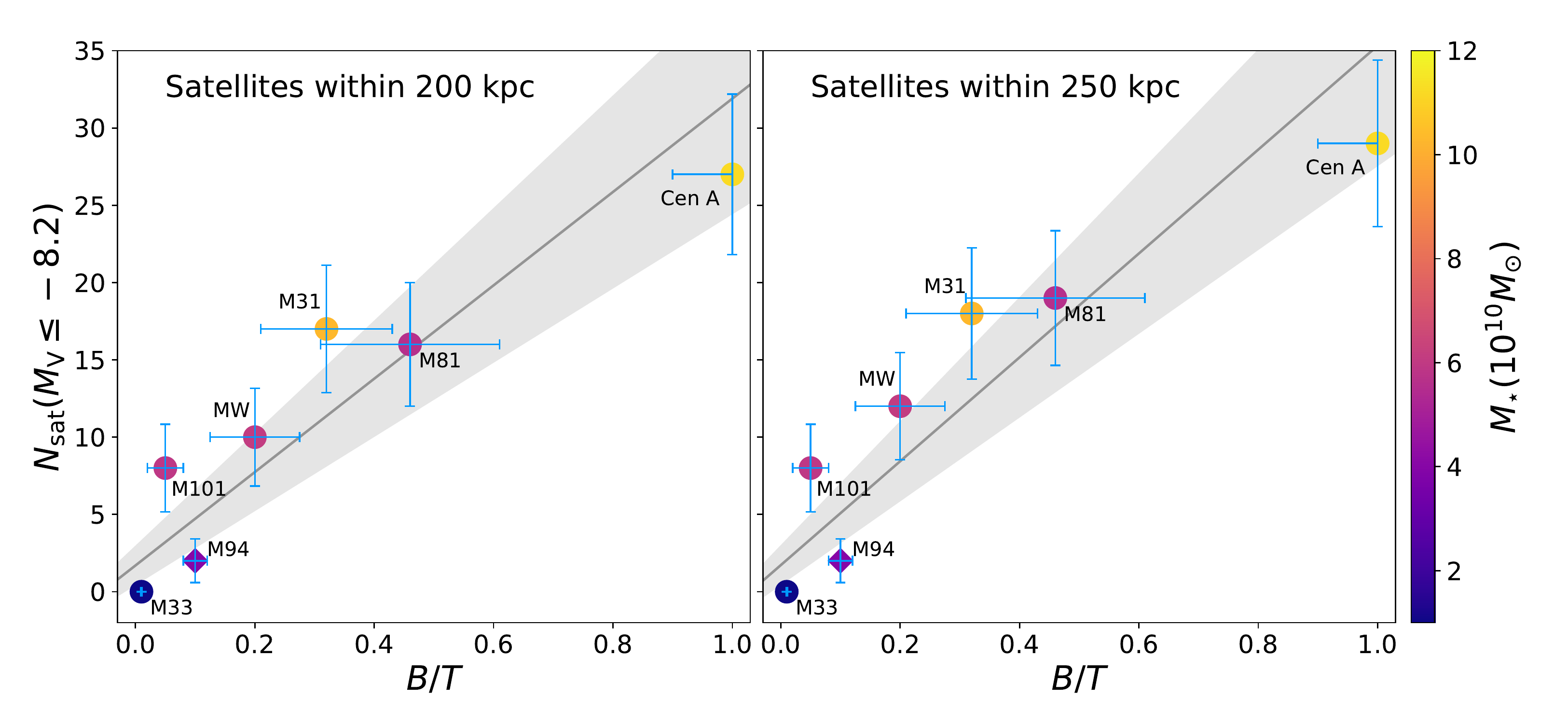}
    \caption{Number of satellites, $N_{sat}$ (with $M_{\rm V}\leq-8.2$) vs. bulge-to-total baryonic mass ratio, $B/T$. The left and right panels correspond to considering satellites within 200 and 250 kpc of their host galaxies, respectively. The color code is $M_{\star}$, the solid grey lines are the results of linear fits taking into account uncertainties on both axes, and the shaded regions are the 1$\sigma$ uncertainty of the fitted results. For both of the analyses, the probability that the correlation arises by chance is only 0.5 percent. Since the current surveys cover only 150 kpc projected distance of M94 (shown by a diamond in this figure), this galaxy is excluded from both the fitting and the correlation analyses. Adding M94 would increase the significance of the correlation.}
    \label{fig:Nsat_BT}
\end{figure*}


\subsection{Bulge-to-total mass ratios}\label{sec:BT}
The information about the $B/T$ ratios of the galaxies in our study are as follows:
\begin{itemize}
    \item For MW, M31, M81, Cen A, and M101, stellar mass and $B/T$ values are directly adopted from \citet{Bell2017}. See their table 1 for a list of references and other measurements.
    \item For M33, using $M_{\rm bulge}=1.14(\pm 0.14) \times 10^{8}M_{\odot}$ reported by \citet{Seigar2011}, and adopting a total mass of $\approx10^{10}M_{\odot}$ reported by \citet{Corbelli2003}, we obtain $B/T=0.01\pm0.001$.
    \item For M94, we adopt the mean (and its standard error) of the $M_{\rm bulge}$ values reported in \citet{Moellenhoff1995} and \citet{Jalocha2010}; $M_{\rm bulge}=4.54(\pm 0.86) \times 10^{9}M_{\odot}$. Considering a stellar mass of $4\times10^{10}M_{\odot}$ \citep{Smercina2018}, we obtain a $B/T=0.1\pm0.002$ for M94.
\end{itemize}
See the second and third columns of Table \ref{tab:data} for a summary.
\begin{table}
    \caption{The data used in this study and their references. See Section \ref{sec:data}.}
    \begin{tabular}{lcccc}
    Galaxy&	$M_{\star}(10^{10}M_{\odot}$) & $B/T$ & $N^{200}_{sat}$ & $N^{250}_{sat}$\\
    \hline
    MW & 6.1 (a) & 0.2$\pm$0.075 (a)& 10 (b) & 12 (b)\\
    M31& 10.3 (a) & 0.32$\pm$0.11 (a)& 17 (c) & 18 (c)\\ 
    M33& 1 (d) & 0.01$\pm$0.001 (d,e)& 0 & 0\\
    M81& 5.6 (a)& 0.46$\pm$0.15 (a)& 16 (f) & 19 (f)\\
    Cen A& 11.2 (a) & $1.0^{+0.0}_{-0.1}$ (a) & 27 (g,h) & 29 (g,h)\\
    M94$^{\dag}$& 4 (i) & 0.1$\pm$0.02 (j,k)& 2$^{\dag}$ (i)& 2$^{\dag}$ (i)\\
    M101 & 5.9 (a)& 0.05$\pm$0.03 (a)& 8 (l,m) & 8 (l,m)\\
    \hline
    \end{tabular}
    (a) \citet{Bell2017}; (b) \citet{McConnachie2012}; (c) \citet{McConnachie_2018}; (d) \cite{Corbelli2003}; (e) \citet{Seigar2011}; (f) \citet{Chiboucas_2013}; (g) \citet{Cronjevic2019}; (h) \citet{Muller2019}; (i) \citet{Smercina2018}; (j) \citet{Moellenhoff1995}; (k) \citet{Jalocha2010}; (l) \citet{Carlsten2019}; (m) \citet{Bennet2019}.\\
     (\dag) Note that the field of M94 is surveyed only out to 150 kpc projected distance, and while we show it in Figure \ref{fig:Nsat_BT}, we do not include it in our correlation analysis.
    \label{tab:data}
\end{table}

\subsection{The correlation}\label{sec:correlation}
The data listed in Table \ref{tab:data} are visualized in Figure \ref{fig:Nsat_BT} by plotting $N_{\rm sat}$ vs. $B/T$. Each galaxy is labeled on this figure and the color code represents the $M_{\star}$. The uncertainties on $B/T$ are values from Table \ref{tab:data}, and we consider an uncertainty of $\pm\sqrt{N_{\rm sat}}$ for the number of satellites. The left and right panels correspond to considering satellites within 200 and 250 kpc of their host galaxies, respectively. The solid grey lines are the result of linear least squared fits to the data: $N^{200}_{sat}=30.2(\pm6.2)B/T + 1.7(\pm1.3)$, and $N^{250}_{sat}=33.6(\pm6.5)B/T + 1.7(\pm1.3)$. In the fitting procedure, we take both $N_{\rm sat}$ and $B/T$ uncertainties into account. The shaded regions reflect the 1$\sigma$ uncertainty of the fitting results. However, we note that we do not aim to present a formula for $N_{\rm sat}$ vs. $B/T$ and the purpose of the fitting is to better see the trend presented by these data.

Our main finding is that using these data we measure the linear correlation coefficient for $N_{\rm sat}$ vs. $B/T$ to be $r=0.94$ and the probability that this correlation arises by chance to have a $p-value=0.005$, for both of the projected distance conditions\footnote{  The $r$ and the $p$ values for the two projected distance conditions differ only after the third decimal places, not reported here.}. In other words, the data yield a 99.5 percent (around $3\sigma$) significant correlation between $N_{\rm sat}$ and $B/T$. We note that M94 is excluded from both the linear fitting and the correlation measurement.

\section{Discussion and Concluding Remarks}\label{sec:discussion}
Figure \ref{fig:Nsat_BT} can be compared with the right panel of figure 4 in \citet{Javanmardi19} which shows $N_{\rm sat}$ vs. $B/T$ for a sample of more than 6000 disk galaxies with stellar masses between 1.2 and 26.3 $\times10^{10}M_{\odot}$ from the Millennium-II simulation \citep{Boylan-Kolchin2009}. This mass range almost encompasses that of the spiral galaxies in our study. \citet{Javanmardi19} measure a linear correlation coefficient of only $r=0.13$ for $N_{\rm sat}$ vs. $B/T$, being consistent with no correlation between these two quantities in the $\Lambda$CDM model.

This can be understood also qualitatively by noting that in the $\Lambda$CDM model, $N_{\rm sat}$ is directly related to dark matter halo mass; the heavier the halo, the larger the number of accreted subhalos \citep[top left panel of figure 3 in][]{Javanmardi19}, and heavier halos accrete also more baryonic mass forming galaxies with statistically larger stellar masses. In addition, in this model, galaxies with larger stellar masses are found to be more likely to grow heavier bulges \citep[figure 2 in][]{Javanmardi19}. Therefore, in the $\Lambda$CDM model, $N_{\rm sat}$ is weakly correlated with $M_{\rm bulge}$, but not with $B/T$. This certainly requires further detailed studies of bulge formation  and its possible connection with large-scale environment in this model (\citeauthor{Romano-Diaz2017} \citeyear{Romano-Diaz2017}, \citeauthor{Tavasoli} in prep.).

An interesting point in the current data is that while M81, M101, and the MW have very similar stellar masses, they have different $N_{\rm sat}$ and very different $B/T$ ratios, implying a lack of correlation between these two quantities and stellar mass.

We note that Cen A has a few features which makes it distinct with respect to the other galaxies in our analysis. It is not a disk-galaxy (with $B/T=1.0$), it has an active nucleus, and it has a perturbed structure. Assuming that these have a significant impact on its satellite population, remeasuring the $N_{\rm sat}$ vs. $B/T$ correlation without Cen A gives $r=0.93$ and $p-value=0.02$, not changing our main result.

In addition, it should be emphasised that as mentioned earlier, M94 is not included in our correlation analysis because its field is surveyed out to only 150 kpc projected distance. Assuming that future surveys do not change the number of satellites of M94 significantly, adding it to the analysis would increase the significant of the correlation to well above the $3\sigma$ confidence level. Actually, even if future surveys increase the number of confirmed satellites of M94 by a factor of 3 or 4, the significance of the $N_{\rm sat}$ vs. $B/T$ correlation would increase even further. It is therefore very important to conduct deeper and wider surveys in the field of this particular galaxy.

In a very recent study, \citet{Carlsten2019b} have reported the detection of satellite galaxy candidates around 10 galaxies including M104 (a.k.a. the Sombrero galaxy, at a distance of around 9.5 Mpc). They report in total 27 satellite candidates within 150 kpc projected distance of this galaxy\footnote{Also confirming the low surface brightness galaxies found by the Dwarf Galaxy Survey with Amateur Telescopes \citep[DGSAT,][]{Javanmardi2016} within 80 kpc projected distance of M104.}. With $B/T>0.7$ \citep{Gadotti2012}, this galaxy can add another point to the large $B/T$ values of the $N_{\rm sat}$ vs. $B/T$ relation. \citet{Carlsten2019b} also report in total 21 satellite candidates for NGC 4565 that has a $B/T=0.25$ \citep{Bell2017}. However, follow up distance measurements for their satellite candidates are essential before including these two galaxies in the analysis. 

As a final remark, it is also worth noting other reported interesting properties of the non-LG galaxies in our study. \citet{Chiboucas_2013} found that M81 has a disk of satellites similar to those of MW \citep{Kroupa2005,Pawlowski2013} and M31 \citep{Metz2007,Hammer2013,Ibata2014}. In addition, \citet{Mueller2018} reported a similar structure around the Cen A galaxy. \citet{Smercina2018} and \citet{Bennet2019} conclude that M94 and M101 have sparse satellite populations, not fulfilling the expectations from the $\Lambda$CDM model. \citet{Bennet2019} suggests a link between the properties of these galaxy groups and their environments reporting that galaxies in more tidally active environments tend to have a larger $N_{\rm sat}$ (see their figure 8).

The latter point is in line with an alternative scenario by \citet{Kroupa2015} arguing that if most of the observed satellites are ancient tidal dwarf galaxies, then a correlation between $N_{\rm sat}$ and $B/T$ can emerge because bulge growth is enhanced significantly in galaxy-galaxy encounters. This would also explain the observed preferred distribution of satellites in rotating disks \citep{Pawlowski2013b,kroupa2012}. This scenario also requires further detailed and quantitative studies \citep[see e.g.][]{Combes2016,Bilek2018,Banik2018}.

Whatever the correct underlying theory will turn out to be, and while further observational studies of these and other galaxy groups are necessary, the main finding of our study is that the current data show a strong and significant correlation between $N_{\rm sat}$ and $B/T$ in the most well-studied galaxy groups in the Local Volume.

\section*{Acknowledgements}
We thank the anonymous referee for the constructive comments. The research leading to these results has received funding from the European Research  Council  (ERC)  under  the  European  Union's  Horizon  2020  research and innovation program (grant agreement No. 695099).




\bibliographystyle{mnras}
\bibliography{references} 


\bsp	
\label{lastpage}
\end{document}